\documentclass[fleqn,usenatbib]{mnras}

\usepackage{mathptmx}
\usepackage[T1]{fontenc}
\usepackage{ae,aecompl}

\usepackage{graphicx}	% Including figure files
\usepackage{amsmath}	% Advanced maths commands
\usepackage{amssymb}	% Extra maths symbols

\title[Quenching history of Cluster Galaxies]{Tracing the Quenching History of Cluster Galaxies in the EAGLE Simulation}

\author[D. Pallero et al.]{
Diego Pallero$^{1}$\thanks{E-mail:dpallero@dfuls.cl},
Facundo A. G\'omez$^{1,2}$,
Nelson D. Padilla$^{3,4}$,
S. Torres-Flores$^{1}$,\newauthor
~R. Demarco$^{5}$,
P. Cerulo$^{5}$ and 
D. Olave-Rojas$^{6}$
\vspace{0.2cm}\\
\\
$^{1}$Departamento de F\'isica y Astronom\'ia, Universidad de La Serena, Av. Juan Cisternas 1200 Norte, La Serena, Chile\\
$^{2}$Instituto de Investigaci\'on Multidisciplinar en Ciencia y Tecnolog\'ia, Universidad de La Serena, Ra\'ul Bitr\'an 1305, La Serena, Chile\\
$^{3}$Instituto de Astrof\'isica, Pontificia Universidad Cat\'olica de Chile, Santiago, Chile \\
$^{4}$Centro de Astro-Ingenier\'ia, Pontificia Universidad Cat\'olica de Chile, Santiago, Chile\\
$^{5}$Departamento de Astronom\'ia, Facultad de Ciencias F\'isicas y Matem\'aticas, Universidad de Concepci\'on, Concepci\'on, Chile\\
$^{6}$Departamento de F\'isica, Facultad de Ciencias, Universidad del B\'io B\'io, Collao 1202, Concepci\'on, Chile\\
}

\date{Accepted XXX. Received YYY; in original form ZZZ}

\pubyear{2015}

\begin{document}
\label{firstpage}
\pagerange{\pageref{firstpage}--\pageref{lastpage}}
\maketitle

\begin{abstract}
We use the EAGLE hydrodynamical simulation to trace the quenching history of galaxies in its 10 
most massive clusters. We use two  criteria to identify  moments when 
galaxies suffer significant changes in their star formation activity: {\it i)} the 
instantaneous star formation rate (SFR) strongest drop, $\Gamma_{\rm SFR}^{\rm SD}$, and {\it ii)} 
a ``quenching" criterion based on a minimum threshold for the specific SFR of $\lesssim$ 10$^{-11}\rm yr^{-1}$. 
We find that a large fraction of 
galaxies ($\gtrsim 60\%$) suffer their $\Gamma_{\rm SFR}^{\rm SD}$ outside the cluster's R$_{200}$. 
This ``pre-processed" 
population is dominated by galaxies that are either low mass and centrals or inhabit low mass hosts 
($10^{10.5}$M$_{\odot} \lesssim$ M$_{\rm host} \lesssim 10^{11.0}$M$_{\odot}$). The host mass 
distribution is bimodal, and galaxies that suffered their $\Gamma_{\rm SFR}^{\rm SD}$ 
in massive hosts ($10^{13.5}\rm M_{\odot} \lesssim M_{host} \lesssim 10^{14.0}M_{\odot}$) 
are mainly processed within the clusters. Pre-processing mainly limits  
the total stellar mass with which galaxies arrive in the clusters. Regarding quenching,  
galaxies preferentially reach this state in high-mass halos 
($10^{13.5}\rm M_{\odot} \lesssim M_{host} \lesssim 10^{14.5}M_{\odot}$). The small fraction of 
galaxies that reach the cluster already quenched has also been pre-processed, linking both criteria as 
different stages in the quenching process of those galaxies. For the $z=0$ satellite populations, 
we find a sharp rise in  the fraction of quenched satellites at the time of first infall, 
highlighting the role played by the dense cluster
environment. Interestingly, the fraction of pre-quenched galaxies rises with final cluster mass. 
This is a direct consequence of the hierarchical cosmological model used in these simulations.

\end{abstract}


%%%%%%%%%%%%%%%%%%%%%%%%%%%%%%%%%%%%%%%%%%%%%%%%%%

%%%%%%%%%%%%%%%%% BODY OF PAPER %%%%%%%%%%%%%%%%%%

\section{Introduction}
\label{sec:intro}

Since the first half of the twentieth century, it has been known that colors reflect the predominant stellar populations in galaxies and that they are related to their morphology \citep{Morgan57}. The colour-morphology \citep{Roberts94} and  color-magnitude relations \citep[]{Chester64, Faber73} are now widely used to study the properties of galaxies. As a result, rather than selecting objects according to their \textit{early-} or \textit{late-type} morphology, galaxies can be separated between \textit{red} and \textit{blue}, which naturally relates with their star formation and metal-enrichment history. Studies in the local universe show that, in general, galaxies present a strong bimodal color distribution \citep[]{Strateva01, Baldry06,Cassata08}, regardless of the environment in which they reside \citep[]{Hogg04, Baldry06}. 
%This suggests different mechanisms driving the evolution of galaxies on each of these regimes \citep[]{Menci05, DeLucia07}. 
Reproducing this bimodality, and understanding the role played by the environment, has become an important goal for galaxy-evolution theories \citep[]{Trayford15, Nelson18}.

%the understanding of different processes that lead to these two sequences, as well as the moment of its formation has been under discussion in the last decade.\\
%In this context, the environment can play an important role in the evolution of galaxies.
%Galaxy clusters present an unique collection of galaxies subjected to various changes due to environmental effects \citep{Treu03}.
One of the first indications that the environment plays a fundamental role in driving the evolution of galaxies was the morphology-density relation \citep[]{Dressler80, Dressler84}. Observational studies have shown that in high density environments there is a greater fraction of galaxies with early-type morphology than in low density environments, and that the fraction of early-type galaxies in clusters rises toward the cluster's center \citep[]{Brough17, Cava17}.

In addition, several studies during the last decades have shown that dense environments can also affect the star formation history of galaxies \citep[]{Gunn72, Dressler80, Moore96, Poggianti01,Boselli05}.
Naturally, the cores of galaxy clusters are an ideal laboratory to study how the environment affects the evolution of galaxies in dense regions and at different redshifts \citep[]{Cayatte90, Smail97, BravoAlfaro00,Boselli05}. Evidence of global transformations for galaxies over time is given by the increasing fraction of spiral in clusters up to z $\sim$ 0.5 \citep[]{Dressler97, Fasano00, Desai07}, and thanks to the fact that high-z clusters are observed to contain more star-forming galaxies compared to present-day \citep[]{Butcher84, Poggianti06}. 

It is also well known that there are differences between the properties 
of galaxies located in the inner and outer regions of galaxy clusters. Some authors 
\citep[]{Kodama01, Treu03}, suggest that this is the result of a variety of 
mechanisms that act at different distances from the cluster center, driving galaxy evolution with different timescales. Moreover, it has been observed (e.g. \citealt{ Dressler13, Hou14, Bianconi18}) that in the outskirts of clusters infalling galaxy partly distribute in the form of groups. 

Theoretical works (e.g. \citealt{McGee09, deLucia12}) suggest that $\sim 25-40\%$ of galaxies belonging to a massive cluster (M$_{\rm halo}~\sim10^{14.5}-10^{15.0}$[M$_{\odot}]$) at z $=$ 0 have been accreted in such groups. %In this sense, according to the hierarchical scenario, galaxy groups are thought to be the buildings blocks of today's rich clusters. 
For the aforementioned reasons, the study of galaxy properties in cluster outskirts, where these systems are still assembling, has attracted the interest of several astronomers (e.g. \citealt{Just10, Cybulski14, Jaffe16}). In particular, some authors (e.g. \citealt{Hou14, Haines15, Bianconi18}) have studied the variation of the fraction of quiescent galaxies with the projected distance from the centers of clusters, finding that out to 3$R_{200}$ clusters are richer in passive galaxies than the field. These results can only be explained if star formation was quenched in galaxies prior to their accretion on to clusters, when they were still members of in-falling groups (pre-processing \citealt[]{Zabludoff98,Fujita04}). The fact that galaxies are ``preprocessed'' in groups before their accretion on to clusters shows that groups of galaxies constitute an important piece in the physics of galaxy formation and evolution, because the processes that take place within them may significantly alter star formation and change the structural and chemical properties of galaxies. Groups provide, then, further laboratories to study the environmental drivers of galaxy evolution (e.g. \citealt{Dressler13,Bianconi18,Olave18}).

%It is also well known that galaxies can strongly interact with the environment since the outskirts of clusters through a variety of different processes \citep[]{Kodama01, Treu03}. Moreover, infalling galaxies are usually \green{est\'as seguro que es usual que las galaxias caen en c\'umulos como miembros de grupos? Cu\'antas galaxias de las que estudias en este art\'iculo caen en los c\'umulos como miembros de halos con m\'as que una galaxia?} observed to belong to  infalling groups which, according to the hierarchical scenario, are thought to be the buildings blocks of today's rich  clusters. Thereby, galaxy groups can constitute a fundamental stage of galaxy evolution thanks to the  "pre-processing" of their members before accretion into the main final cluster. These groups provide another suitable laboratory to study the physical processes that alter the morphology and other observable properties of galaxies, such as  color and star-formation rate \citep{Fujita04}. 

\begin{figure*}
    \centering
    \includegraphics[width=\textwidth]{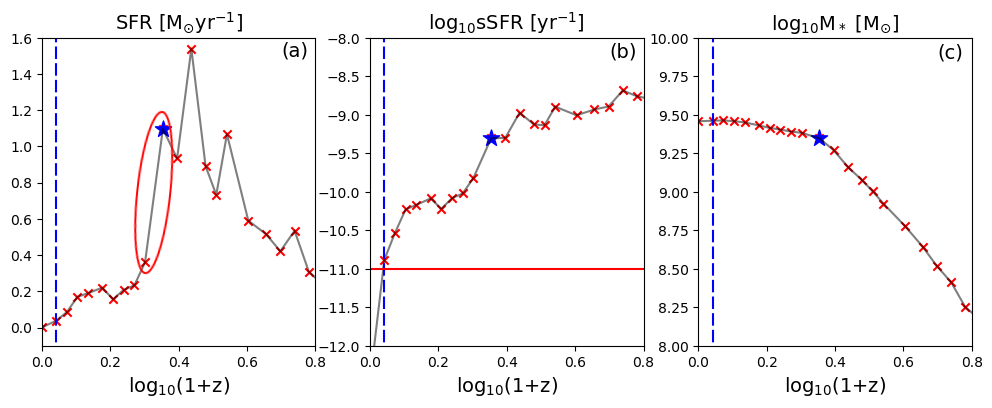}
    \caption{Examples of the selection criteria used in this work to determine whether a galaxy was processed or quenched, for a random galaxy in our sample. Panel (a) shows the star formation rate against redshift. The red ellipse highlights the strongest drop in the star formation activity, and the blue dashed line corresponds to the time of the first infall into the final cluster's R$_{200}$. Panel (b) shows the specific star formation rate of the galaxy against redshift. The red solid line shows the critical star formation rate imposed by our selection criterion to define quenched galaxies. The blue dashed line corresponds to the first infall into R$_{200}$. Panel (c) shows the growth of the galaxy stellar mass through cosmic time. The blue star indicates the moment when the processing started. We can see that the growth of the stellar mass is suppressed after the strongest drop, and that the specific star formation rate decreases abruptly after passing  R$_{200}$.}
    \label{fig:criterias}
\end{figure*}

At z $\sim$ 0,  the specific star-formation rate of galaxies in dense environments is significantly smaller than in lower density regions \citep[]{Hashimoto98,Lewis02,Kauffmann04, Gray04,Balogh07}. Additionally, higher fractions of quiescent or passive galaxies are found in dense regions \citep[]{Poggianti99, Baldry06, vandenBosch08, Gavazzi10, Haines13}. These studies also provide evidence that the star-forming activity and galaxy morphology can be correlated with z $\sim$ 0 galactic stellar mass. Less massive galaxies also are more susceptible to environmental effects, indicating that the quenching of star formation can be accelerated in dense environments\citep[]{deLucia12, Muzzin12, Jaffe16}.

In spite of the observed relation between environment and cessation of star formation activity, 
i.e. \textit{``environmental quenching"}, it is important to take into account internal  
process that can drive galaxy quenching. This process, known as \textit{``mass quenching"} or \textit{``internal quenching"}, can arise as a result of, e.g. internal gas consumption, supernova and AGN feedback, star formation feedback or halo gas heating (see e.g.\ \citealt{Peng10},  \citealt{Efstathiou_2000}, \citealt{Croton_2006}, \citealt{Dekel_2008}, \citealt{Cantalupo_2010}). The dominance of one way over the other is where the dichotomy of \textit{``nature versus nurture"} was born, and has been one of the main subjects of study for extragalactic astronomy in the last years.

According to \citet{Oesch16}, quenching may start shortly after the first appearance of the galaxies, at roughly z $\sim$ 11, but the environment does not play an important role until z $ \sim $ 1.6, with the environmental quenching efficiency rising by a factor of $ \sim $ 3.5 between z $ \sim$ 1.6 and z $\sim$ 0.9. The quenching efficiency is defined as the ratio between passive galaxies in clusters and passive galaxies in the field, quantifying the number of galaxies that would be star-forming if they were in the field \citep[]{Peng10,Peng12,Nantais16}. Nevertheless, a study of the sSFR and the fraction of star-forming galaxies in clusters at z $\sim$ 1 from the GCLASS survey \citep{Muzzin12} shows that  mass quenching dominates over environmental quenching, at least at this redshift. \citet{Balogh16}, using another cluster sample from GCLASS, found that the mechanisms driving the quenching at z $ > $ 1 may be different from those at z $\sim$ 0. On the one hand, at high redshift, the cessation of star formation is mainly driven by a combination of gas consumption (due to an enhancement of star formation) and gas outflows as a result of supernovae and AGN feedback. On the other hand, at low redshift, dynamical mass removal mechanisms (due to environment) may be the main driver for the quenching of galaxies in clusters.

A detailed description of the main mechanisms that lead to environmental quenching is provided by \citet{Boselli06} and \citet{Jaffe16}. They   separate these mechanisms in three broad categories: 
\begin{itemize}
\item Gravitational interactions between galaxies: Mergers can change drastically the star formation history of galaxies, as well as their morphology and kinematics. This phenomenon is usually observed in low-density environment such as groups of galaxies. In high-density environments, galaxies can experience harassment from other cluster members, through fast and aggressive encounters
\citep[]{Toomre72,Barnes96,Walker96,Moore99}; 

\item Interactions between galaxies and the intra-cluster medium: Ram-pressure from the intracluster medium can strip the gas of the galaxies and remove their interstellar medium
\citep[]{Gunn72, Abadi99, Quilis00, Vollmer01,Jaffe15, Benitez13}; 

\item Gravitational interactions between clusters and galaxies: The tremendous gravitational potential of the cluster can perturb some observable properties of the members, inducing gas inflows, forming bars, compressing the gas or concentrating the star formation \citep[]{Miller86, Byrd90, Boselli06}.

\end{itemize}

%n one hand, strong merger of disk galaxies, can produce a massive elliptical galaxy (\cite{Toomre72}, \cite{Barnes96}), and on the other hand, accretion of satellite galaxies for a disk galaxy, can produce an S0 galaxy (\cite{Walker96}).
%It should be noted, that galaxy-galaxy interaction are uncommon in galaxy cluster due to their high dispersion velocities, so this kind of interaction is usually observed in low mass galaxy-groups, in high-density environment, galaxies experience fast and aggressive encounters with another cluster members, these process is known as harassment (\cite{Moore99}), and can strip the HI from galaxies leading to a cessation of star formation    (\cite{Duc08}), this effect, is more severe for low surface galaxies toward the cluster-cores(\cite{Moore99}, \cite{Smith15}).

%Ram-pressure, can eject the gas from the galaxies and remove the interstellar medium (ISM) from these, when it passes through dense regions 
 
%The efficiency of the stripping depends on both, the density of the ICM and the square of the galaxy's dispersion velocity. Thus, when the ram-pressure is stronger than the gravitational interaction between the galaxy an their ISM, the diffuse component will be stripped 
%(luego ver si terminas la idea o lo dejas ahí.)

%The tremendous gravitational potential of the cluster can perturb some observable properties of the members inducing for example gas inflows, forming bars, compressing the gas or concentrating the star formation, nevertheless, gas hardly is directly removed in this kind of interaction 

It is precisely because of this complex nature of the environmental quenching that it is difficult to separate the aforementioned processes. It is expected that, at least, some of these processes act simultaneously and that they are effective in different overlapping regions of the cluster. Some studies show that the effectiveness of these processes is linked to the galaxy's cluster-centric distance \citep{Moran07}.

\begin{figure*}
\centering
\includegraphics[width=\textwidth]{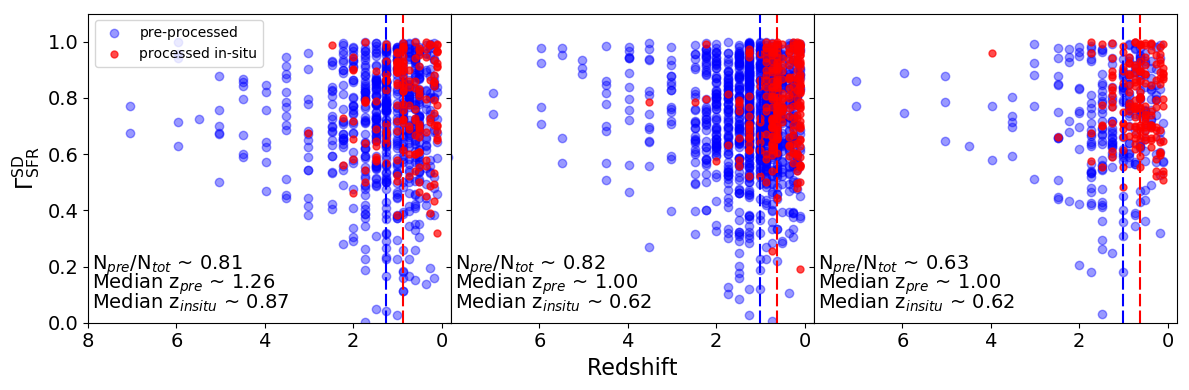}
\caption{ Distribution of $\Gamma_{\rm SFR}^{\rm SD}$ (normalized instantaneous strongest drop of the SF activity), as a function of the redshift at which it takes place. Blue dots correspond to galaxies that suffer their $\Gamma_{\rm SFR}^{\rm SD}$ outside the cluster's R$_{200}$ (pre-processed), while red dots to those that suffer their $\Gamma_{\rm SFR}^{\rm SD}$ inside R$_{200}$ (processed in-situ). The medians for both samples are indicated by the dashed lines. The panels are organized from left to right as high-mass clusters ($14.6<\log_{10} {\rm M}_{\rm host}~[\rm{M}_{\odot}]<14.8$), intermediate-mass clusters ($14.3<\log_{10} {\rm M}_{\rm host}~[\rm{M}_{\odot}]<14.6$) and low-mass clusters ($14.0<\log_{10} {\rm M}_{\rm host}~[\rm{M}_{\odot}]<14.3$), respectively.}
\label{fig:sfr_redshift}
\end{figure*}

A good approach to study the mechanisms that impact galaxy evolution is through cosmological models \citep[]{Fujita04, Wetzel13, Vijayaraghavan13, Schaye15, Nelson18}. Several works have used simulations to understand the properties of galaxies in different environments, and how their evolutionary history results in changes of their properties such as colors, stellar mass and star formation rate \citep[]{Trayford15, Trayford16,Katsianis17, Tescari18, Nelson18}. Hydrodynamical simulations can be used to define and test different criteria that can be used to understand the processes that drive galaxies to be quenched. Simulations also allow us to follow the evolution of galaxies in different environments and the evolution of their properties from z $\sim$ 20 to z = 0. Since clusters at z $>1.5$ are difficult to detect, due to the fact that they are still in an assembling process, simulations are a helpful tool to study the role that the environment plays at such high redshift (see e.g. \citealt{Overzier_2016}).

In this paper, we use the public database from the state-of-the-art EAGLE hydrodynamic simulations \citep{Schaye15, Crain15, McAlpine16} to 
trace the evolution history of the satellite galaxies that belong to the ten most massive clusters at z $\sim$ 0. 
We aim at identifying the environment in which galaxies preferentially cease their star formation and 
signatures that could be used to determine the main physical mechanism leading to the cessation of star formation of cluster satellite galaxies. We compare the results obtained from two different criteria to identify when star formation in galaxies significantly drops. The hydrodynamic simulations of the EAGLE project are perfectly suited for this study since they provide the possibility to study the evolution of galaxies and their properties.

This paper is organized as follows: in Section 2 we describe the EAGLE simulation, its main characteristics and the main potentialities that it provides for this study. In Section 3 we define the two criteria used in this work to locate those moments when galaxies suffer an important variation in their star formation; in Section 4 we describe the results obtained using our two approaches, putting special interest on the environment where these events take place. Finally, in Section 5 we summarize our main conclusions and compare our results with both observational and theoretical works. A brief discussion of some future projects are presented in this section as well.

%study field galaxies and galaxies in groups near the neighborhoods of clusters of galaxies of $\sim 10 ^{14} $ solar masses, 

%whose star formation ceased before infall to the cluster, and we will call "processed in-situ" to the galaxies whose star formation was ceased inside the cluster.
%looking for pre-processing and quenching takes place before a galaxy falls into a cluster, or if it takes place within the cluster.  We will study the extent out to which the environment of the cluster contributes to the quenching of star formation.

%In order to develop this project, we need an appropriate sample to study the evolutionary history of the galaxies and this particular simulation provides us an hydrodynamical high-resolution model, with a size-box big enough to carry this study.
%We also need an appropriate definition of quenching and pre-processing to lead our study. We will define quenching as the moment on the history of the galaxy where presents its worst infall of specific star formation rate (sSFR).
%Our final aim is to use this study to select pre-processed galaxies in EAGLE, and to find a correlation between this process and resolved properties of these galaxies, to be able to devise a technique that will allow their detection in observed clusters using IFU observations.

\section{THE EAGLE SIMULATION}

The EAGLE project, is a suite of cosmological hydrodynamical N-body simulations. These simulations were run with a modified version of the GADGET-3 code, wich is an improved version of GADGET-2 \citep{Springel05}. All the simulations adopt a flat $\Lambda$CDM cosmology whose parameters were calibrated with the data obtained by the Planck mission \citep{Planck14}; $\Omega_{\Lambda}$ = 0.693, $\Omega_m$ = 0.307, $\Omega_b$ = 0.04825, $\sigma_8$ = 0.8288, Y = 0.248 and H$_{0}$ = 67.77 km s$^{-1}$.\\
In particular, for this work we select our sample of galaxies from the main simulation, referred to as L100N1504, which consists of a periodic box with a volume of (100cMpc)$^3$, initially containing 1,504$^3$ gas particles with an initial mass of 1.81 $ \times 10^6$M$_\odot$, and the same amount of dark matter particles with a mass of 9.70 $\times 10^6$M$_\odot$.\\
Each simulation counts with 29 discrete snapshots from redshift 20 to 0, with a time span between consecutive snapshots ranging from 0.3 to 1Gyr. Radiative cooling and photoheating are implemented following \cite{Wiersma09}, assuming an optically thin X-Ray/UV background \citep{Haardt01}. Star formation is implemented stochastically following \cite{Schaye08}, and using the metallicity-dependent density threshold shown in \cite{Schaye04}. This reproduces the observed Kennicutt-Schmith law \citep{Kennicutt98}. Each particle is assumed to be a single-age stellar population, with a Chabrier initial mass function  in the range 0.1 M$_\odot$ - 100M$_\odot$ \citep{Chabrier03}.

Stellar evolution is modelled as shown in \cite{Wiersma09b}, and chemical enrichment is followed for the 11 elements that most contribute to radiative cooling from massive stars (Type II supernovae and stellar winds) and intermediate-mass stars (Type Ia supernovae and AGB stars). Following \cite{Vecchia12}, the thermal-energy product of stellar feedback is stochastically distributed among the gas particles surrounding the event without a preferential direction.

The EAGLE project calibrated the free parameters associated with stellar feedback to match the observations for the stellar mass function in a range of 10$^{8}$M$_\odot$ - 10$^{11}$M$_\odot$ and the size-mass relation for galaxies in a range of 10$^{9}$M$_\odot$ - 10$^{11}$M$_\odot$ \citep[]{Schaye15,Furlong15,Furlong17}.
The appropriate calibration of the subgrid physics and the good agreement with the observational data make these simulations our best tool to study the evolution in the star formation of galaxies in these mass ranges for different environments.

The halo catalogues provided in the public database (used in this work) were built using a friend-of-friends (FoF) algorithm which identifies dark matter overdensities following \cite{Davis85}, considering a linking length of 0.2 times the average inter-particle spacing. Baryonic particles are assigned to the FoF halo of their closest dark matter particle. Subhalo catalogues were built using the \textsc{subfind} algorithm \citep[]{Springel01,Dolag09}, which identifies local overdensities using a binding energy criterion for particles within a FoF halo. We will define as galaxies those structures recognized by the \textsc{subfind} algorithm \citep[]{Springel01,Dolag09} which posses a total stellar mass greater than 10$^{8}$ M$_\odot$ and a total mass greater than 10$^{9}$ M$_\odot$. These masses are obtained by direct summation of the corresponding particles; i.e., particles bound to the subhalo according to the \textsc{subfind}. Since the simulations considered in this work have a baryonic mass resolution of $1.6 \times 10^6$ M$_\odot$ and dark matter mass resolution of $9.1 \times 10^6$M$_\odot$, we ensure that we have at least 100 baryonic and dark matter particles in each galaxy, thus avoiding spurious results and non-physical detections. 
The analyzed clusters correspond to the 10 most massive in the simulation at $z = 0$. They all posses M$_{200}$  > 10$^{14}$ M${_\odot}$. A galaxy is defined as satellite if it can be found inside the host R$_{200}$ at $z = 0$.

\section{The end of the star forming phase: Definitions}

According to \cite{Peng10}, the quenching of a galaxy is the result of a process with two different components. A continuous component associated with internal galactic processes such as star formation and AGN feedback, and a ``once-only" component due to environmental processes. Note, however, that other mechanisms like mergers may also have an important effect on the star formation activity. 

To determine the moment when the star formation activity in a galaxy drops in a significant way, two different criteria are introduced: one based on the maximum drop of the SFR between two consecutive snapshots of the simulation, and the other based on a minimum threshold for specific star formation rate (sSFR). 
The first criterion seeks to identify those mechanisms that abruptly reduce star formation in galaxies, while the second one is meant to define when a galaxy is actually quenched, that is, it is no longer forming stars (e.g. \citealt{Weinmann10,deLucia12,Wetzel12}). The aim of using these two criteria is to determine and understand the different stages of quenching and how they are affected by the environment. 
From now on, we will refer to a galaxy as ``processed'' when it suffers its strongest drop, whereas we will refer to a galaxy as ``quenched" when it reaches the imposed threshold in sSFR. 

\begin{figure*}
\centering
\includegraphics[width=\textwidth]{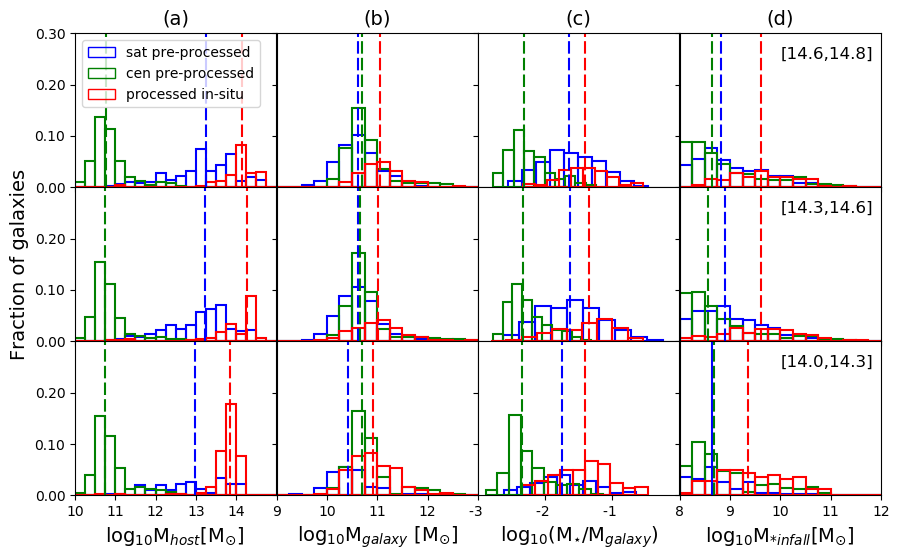}
\caption{Mass distribution of galaxies and their hosts at key moments related to the $\Gamma_{\rm SFR}^{\rm SD}$. Each row shows the results obtained after stacking the distribution of galaxies associated with clusters within different mass ranges. Column (a) shows the mass 
distribution of the
host of each galaxy at the moment of their $\Gamma_{\rm SFR}^{\rm SD}$. Column (b) shows the total mass distribution of the galaxies at their $\Gamma_{\rm SFR}^{\rm SD}$. Column (c) shows the stellar mass fraction distributions at $\Gamma_{\rm SFR}^{\rm SD}$. Column (d) shows the stellar mass distribution of galaxies at the time of their first infall into the cluster they belong to at $z = 0$. Blue, red and greed bars correspond to galaxies pre-processed, in-situ processed, and processed as centrals, respectively. The dashed lines indicate the median of each distribution.}
\label{fig:fraction_strongestdrop}
\end{figure*} 

\begin{figure*}
\centering
\includegraphics[width=\textwidth]{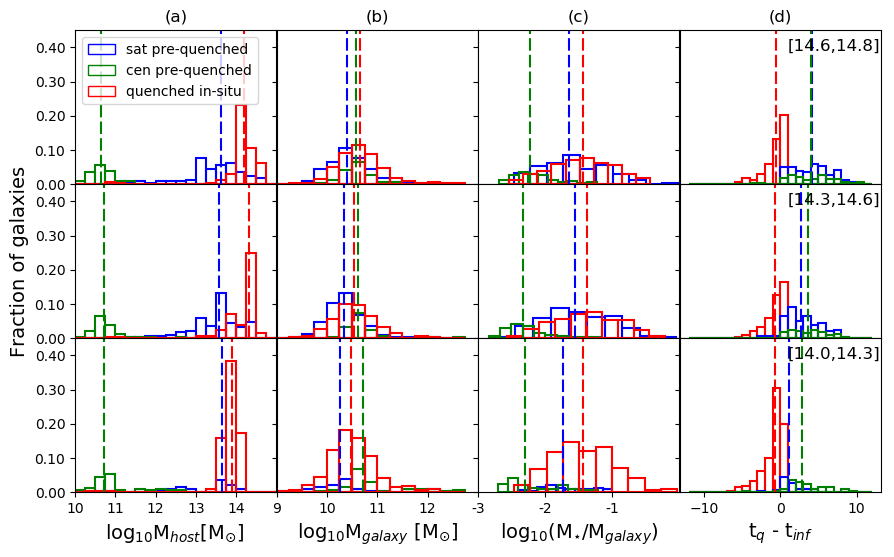}
\caption{Mass distribution of galaxies and their hosts at the moment when they reach their quenching state. Each row shows the results obtained after stacking the distribution of galaxies associated with clusters within different mass ranges. Column (a) shows the mass  distribution of the
host of each galaxy. Column (b) shows distribution of galaxies' total mass. Column (c) shows the stellar mass fraction distributions. Column (d) shows the distribution of times, in lookbacktime, at which galaxies become quenched. Blue, red and greed bars correspond to galaxies pre-quenched, in-situ quenched, and quenched as centrals, respectively. The dashed lines indicate the median of each distribution.}
\label{fig:fraction_wetzel}
\end{figure*}

\subsection{SFR Strongest Drop} 
One of our goals is to identify the mechanisms that can abruptly reduce the star formation in galaxies. For this purpose, we first calculate for each galaxy the variation of the star formation rate between two consecutive snapshots in the simulation, normalized by the star formation in the earliest snapshot. This is

\begin{equation}
    \Gamma_{\rm SFR} = \frac{{\rm SFR}_{i+1} - {\rm SFR}_{i}}{{\rm SFR}_{i}},
\end{equation}
where the subscript $i$ indicates the simulation snapshot, and $i + 1$ is at a lower redshift than $i$.
$\Gamma_{\rm SFR}$ is computed only if the difference between the SFR value in the two snapshots is larger than $1 \times 10^{-3}$M$_\odot$ yr$^{-1}$. This constraint was imposed to avoid measures of $\Gamma_{\rm SFR}$ for galaxies already quenched.
%$\Gamma_{\rm SFR}$ is computed only if the SFR value in both snapshots is larger than $1 \times 10^{-3}$M$_\odot$ yr$^{-1}$. 
f{We then define $\Gamma_{\rm SFR}^{\rm SD}$ as the fraction of star formation lost at the  moment when the strongest drop occurs, i.e. }

\begin{equation}
    \Gamma_{\rm SFR}^{\rm SD} = |\min{\Gamma_{\rm SFR}}|
\end{equation}

We refer to this method as the ``Strongest drop selection criterion''. $\Gamma_{\rm SFR}^{\rm SD}$ takes into account those episodes when a ``once-only" event affects the star formation activity of the galaxies but does not take into account any rejuvenation scenario that could take place afterwards. For this reason, it is not a good tracer of definitive quenching. However, the information gathered by this criterion allows us to find the epochs at which the galaxy suffers a ``processing" event, in particular the most significant one. An example of this selection criterion is shown in Figure \ref{fig:criterias}, panel (a),  where we plot, as a function of time, the star formation rate of a random galaxy in our sample. The red ellipse highlights the moment when $\Gamma_{\rm SFR}^{\rm SD}$ takes place.
In particular for this galaxy, the $\Gamma_{\rm SFR}^{\rm SD}$ is the result of several processes that heat and remove its cold gas content, producing a stagnation in the evolution of the stellar mass and a small decrease in the total gas mass of the galaxy. Unfortunately, we cannot isolate the different mechanisms that produce this processing event due to the lack of temporal resolution. We will further explore this in a future work using a better suited simulation.

\subsection{Critical sSFR criterion} 
\label{sec:def_ssfr}

We further wish to define a criterion that aims at identifying the moment when the galaxies reach a definitive state of ``quenching". Several different definitions of ``quenched galaxy" have been proposed in the literature. Here we used the criterion used in \citet{Wetzel13}. According to this criterion, a galaxy can be considered effectively quenched once it reaches a sSFR$^{\rm Q} = 10^{-11}$ yr$^{-1}$. At this point the galaxy is considered to be passive. From now on we will refer to those galaxies with a sSFR lower than sSFR$^{\rm Q}$ as ``quenched galaxies", and we will call this selection criterion the ``Critical sSFR Selection Criterion''. When using this semi-observational definition, we will only focus on galaxies that are quenched at redshift $z = 0$.  This is to ensure that the selected galaxies will not suffer a rejuvenation process during their evolution. From each of our quenched galaxies, we will extract information about the environment and the time when the quenching state is reached. 

An example of this selection criterion is shown in Figure~\ref{fig:criterias}, panel (b), where the sSFR is shown for the same galaxy from the previous example as a function of time. The red line indicates the sSFR threshold  established in previous works \citep[]{Weinmann10,deLucia12,Wetzel12,Wetzel13} for passive galaxies. In particular for the galaxy shown in the example, the critical star formation is reached once  it crosses the R$_{200}$ of the cluster for the first time, showing the importance of dense environments in the quenching of star formation.

%Here are shown our results obtained using the two criteria aforementioned. The sample was separated among the galaxies that fulfilled the imposed conditions by the criteria before the infalling to the cluster, and those that fulfilled them inside the cluster. The entire study was performed on the 10 most massive clusters in the simulation but, for reading simplicity, only the results obtained for the most massive cluster on the simulation is shown on this article. If the reader is interested in seeing the plots obtained for the others clusters, please refer to the appendix.\\

%\subsection{Strongest drop criteria}
%According to \cite{Peng10}, quenching of galaxies is derived by a continuous component due to the star formation itself and a ``once-only" component due to environmental changes. Also, throughout the life of galaxies, some mechanisms as majors and minor mergers, gas-stripping and the presence of an AGN can dramatically change their sSFR.

\section{Results}

We wish to study the dependencies of star formation quenching on environmental and internal processes focusing on dense environments such as those that can be found in galaxy clusters. For this it is necessary to characterize the properties of individual galaxies such as stellar mass, sSFR and total mass, as well as the 
overall properties of the host cluster such as total mass and virial radius. We will study how these properties evolve as a function of time, and focus on those moments where individual galaxies experience sharp falls in their star formation rates.

As previously discussed in Section~\ref{sec:intro}, in this work we focus on the population
of galaxies associated with the 10 most massive clusters of the EAGLE simulations. In order to study properties of these galaxies as a function of clusters mass with better statistics, the clusters were stacked in three different bins of $z=0$ total mass:
\begin{itemize}
    \item high mass: $14.6<\log_{10} {\rm M}_{\rm host}~[\rm{M}_{\odot}]<14.8$,
    \item intermediate mass: $14.3<\log_{10} {\rm M}_{\rm host}~[\rm{M}_{\odot}]<14.6$,
    \item low mass: $14.0<\log_{10} {\rm M}_{\rm host}~[\rm{M}_{\odot}]<14.3$.
\end{itemize} 
We will refer to these three categories as HMC, IMC and LMC, respectively. The numbers of cluster that fall in each bin are 2 for the HMC, 5 for the IMC and 3 for the LMC. In this section we present our results based on the two previously defined criteria to identify the time at which the star formation activity of a galaxy is significantly altered. 
We will use the terms {\it pre} and {\it in-situ} for galaxies that suffer the previously described processes  inside or outside the cluster R$_{200}$, respectively.

\subsection{Strongest Drop Selection Criterion}
\label{sec:sd}

We first focus on abrupt changes in the SF activity. We start by computing $ \Gamma_{\rm SFR}^{\rm SD}$ for all galaxies that belong to the 10 most massive clusters at $z = 0$. Our goal is to assess  where and when they suffer their most significant processing event. The total number of galaxies in the HMC, IMC and LMC bins are N$_{\rm gal}$ = 846, N$_{\rm gal}$ = 1430 and N$_{\rm gal}$ = 421, respectively. Note that the differences in the number of galaxies is mainly due to the number of clusters that fall in each mass bin. 

In Figure~\ref{fig:sfr_redshift} we show $ \Gamma_{\rm SFR}^{\rm SD}$ for all galaxies as a function of the redshift at which this event takes place. The blue and red dots correspond to pre- and in-situ processed galaxies, respectively. 
The different panels show the results for the different mass bins. We can clearly see that, for the pre-processed population, there is no preferential redshift for $\Gamma_{\rm SFR}^{\rm SD}$ to take place. Note as well that there is no clear correlation between redshift and the typical value of $ \Gamma_{\rm SFR}^{\rm SD}$ for both populations. 
This indicates that these ``once-only" events that significantly affect star formation activity are not associated with any preferential epoch. 

In all mass bins, the majority of the galaxies have been pre-processed. Interestingly, for in-situ processed galaxies, $\Gamma_{\rm SFR}^{\rm SD}$ typically occurs at lower values of redshift than for pre-processed galaxies. This can be seen from the dashed vertical lines, which indicate the median redshift for each population. Note as well that the pre-processed fraction grows with clusters mass, but the median in redshift for pre-processing remains the same regardless of the mass bin. This shows that, although more massive clusters accrete a greater number of pre-processed galaxies, the redshift at which $\Gamma_{\rm SFR}^{\rm SD}$ typically takes place is independent of the $z = 0$ mass of the clusters in which galaxies reside.
 
% \textbf{DIEGO: Esto quien lo sugirio? This result is a consequence of the criterion used to separate the pre-processed from the in-situ galaxies. As the number of galaxies inside the structure rise with radii, at high-redshift the cluster is not big enough to have a representative sample}
 
To understand how the processing affects the evolution of galaxies and which is the role played by the environment, we characterize the mass distribution of the hosts in which these galaxies resided when they  suffered their $\Gamma_{\rm SFR}^{\rm SD}$. In Figure~\ref{fig:fraction_strongestdrop}, panels a), we show the fraction of galaxies per bin of host halo mass, M$_{\rm host}$, at the time of $\Gamma_{\rm SFR}^{\rm SD}$. Fractions are expressed with respect to the total galaxy sample. We split our sample in three populations: pre-processed central galaxies, pre-processed satellites, and in-situ processed satellites. Note that for the pre-processed central population, the mass of the host where the galaxies reside at the moment of processing is nearly the mass of the galaxy itself. From this panel we can clearly see that the three populations are well separated in the host mass distribution, regardless of the mass of the cluster. The median M$_{\rm host}$ of each population is indicated with dashed lines. As we can see, according to the criterion $\Gamma_{\rm SFR}^{\rm SD}$, central pre-processed galaxies tend to suffer their $\Gamma_{\rm SFR}^{\rm SD}$ in low-mass halos, preferentially in halos with total mass  between $10^{10.5}  \lesssim $ M$_{\rm host}$[M$_{\odot}]    \lesssim 10^{11.0}$. For galaxies pre-processed as satellites, $\Gamma_{\rm SFR}^{\rm SD}$ occurs in a large variety of halo masses, ranging between $10^{11}  \lesssim $ M$_{\rm host}$[M$_{\odot}]    \lesssim 10^{13.5}$, with a median near $ 10^{13.0}$M$_{\rm host}$[M$_{\odot}]$ regardless of the mass of the cluster (the typical mass of galaxy groups). On the other hand, for the in-situ processed, it preferentially occurs in higher mass halos, with total masses larger than  $ 10^{14.0}$M$_{\rm host}$[M$_{\odot}]$. 

\begin{figure*}
\centering
\includegraphics[width=\textwidth]{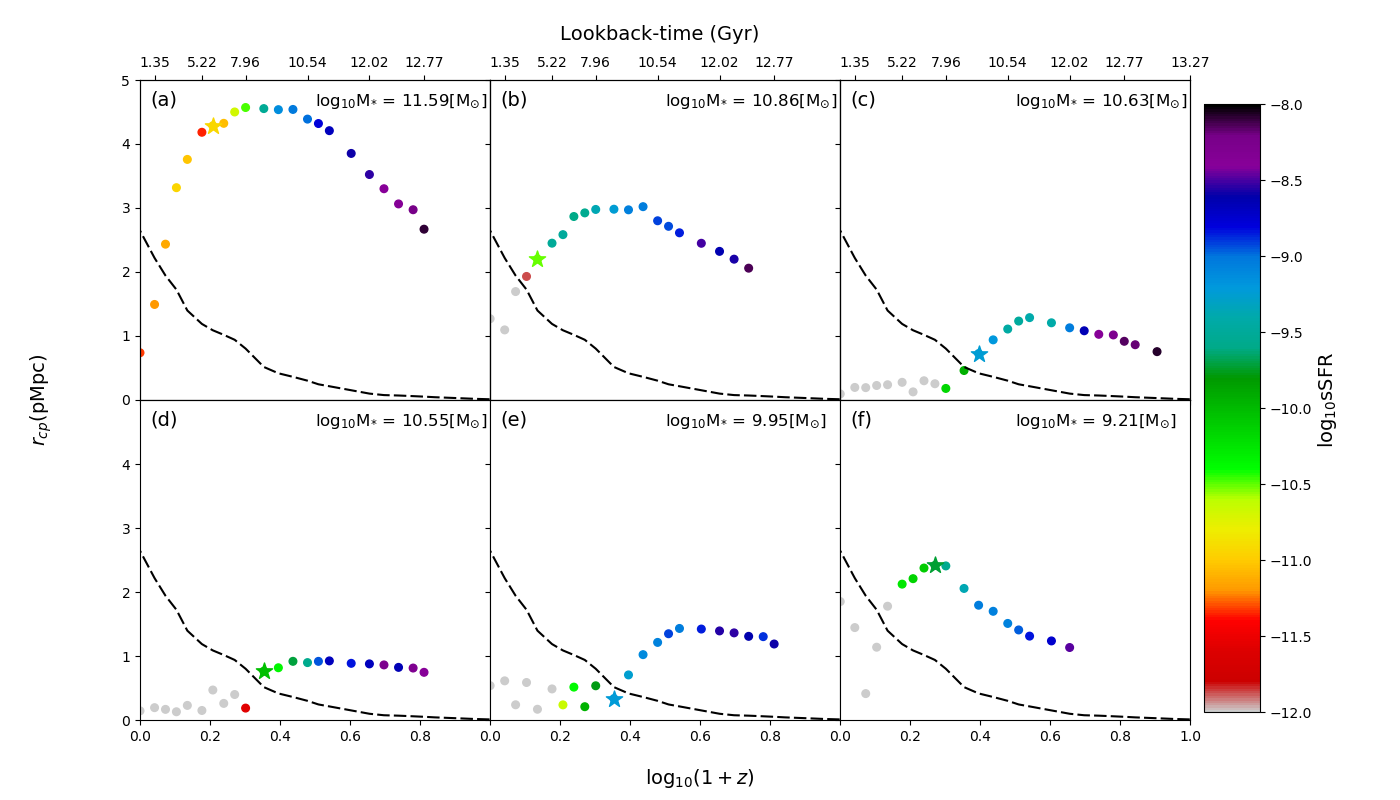}
\caption{Time evolution of cluster-centric distance for a subset of four galaxies in our sample. In panel a) we show a galaxy pre-quenched as central, in panels b) and f) we have galaxies pre-quenched as satellite, and for panels c), d) and e), we have galaxies quenched in-situ. The color coding indicates the sSFR at each time. The dashed line shows the time evolution of the cluster R$_{200}$ and the star shows the moment when the galaxies suffer their ``processing'' event. The label indicates the galaxy stellar mass at $z = 0$.}
\label{fig:story_ssfr}
\end{figure*}

To explore the relation between $\Gamma_{\rm SFR}^{\rm SD}$ and environment we compute, for the overall processed galaxy population, the distribution of total mass (M$_{\rm galaxy}$) and the stellar mass fraction (M$_{\star}$/M$_{\rm galaxy}$) at the time they suffer their $\Gamma_{\rm SFR}^{\rm SD}$. These are shown on panels b) and c) of Figure~\ref{fig:fraction_strongestdrop}, respectively.  
In general we find that in-situ processed galaxies tend to have a  
marginally larger M$_{\rm galaxy}$ than pre-processed galaxies. Interestingly, the difference in 
(M$_{\star}$/M$_{\rm galaxy}$) for these three populations is significantly more evident, with the central pre-processed galaxies showing the lowest stellar mass fractions. This is in agreement with the results shown in Figure~\ref{fig:sfr_redshift}, where we show that $\Gamma_{\rm SFR}^{\rm SD}$ for the in-situ population occurs at lower redshift, thus giving more time to these galaxies to grow in stellar mass. 
Note as well that there is a preference for pre-processing to occur in galaxies when they still remain as centrals, specially for the LMC bin, as shown by the green bars.
We found that, for the pre-processed population, 54.07\% in the HMC bin, 52.14\% in the IMC, and 69.81\% in the LMC were pre-processed as centrals.

As expected for central
galaxies, the M$_{\rm host}$ and  M$_{\rm galaxy}$ distributions are similar. In 
Figure~\ref{fig:fraction_strongestdrop}, panel d), we show the distribution of stellar mass, 
M$_{\star}$, for all galaxies at the time of the first R$_{200}$ crossing. We can clearly see that the difference in M$_{\star}$ between in-situ and pre-processed galaxies is not only present at the time of $\Gamma_{\rm SFR}^{\rm SD}$, but pre-processed galaxies tend to arrive in the cluster with a significantly lower stellar mass. These results
suggest that one of the strongest effects associated with this pre-processing is to limit 
the final stellar mass of satellites in galaxy clusters. As an example, in
Figure~\ref{fig:criterias}, Panel c, we show how the $\Gamma_{\rm SFR}^{\rm SD}$ 
significantly affects the subsequent growth of M$_{\star}$ in a galaxy.
For the pre-processed population, we have derived the time difference between the 
infall time, $t_{\rm inf}$, and the pre-processing time, $t_{\rm proc}$. In general, we find 
that $(t_{\rm proc} - t_{\rm inf})$ is smaller for satellite galaxies than for centrals, 
and that this quantity grows with cluster mass. This result explains the difference in stellar mass ratio at the moment of the processing seen in Figure~\ref{fig:fraction_strongestdrop} for the pre-processed population. Central galaxies suffer their pre-processing earlier than the satellite sample and, despite the fact that both populations shows similar M$_{\star}$ at the moment of the infall, those which had their strongest drop as centrals are more dark matter dominated.

\begin{figure*}
\centering
\includegraphics[width=\textwidth]{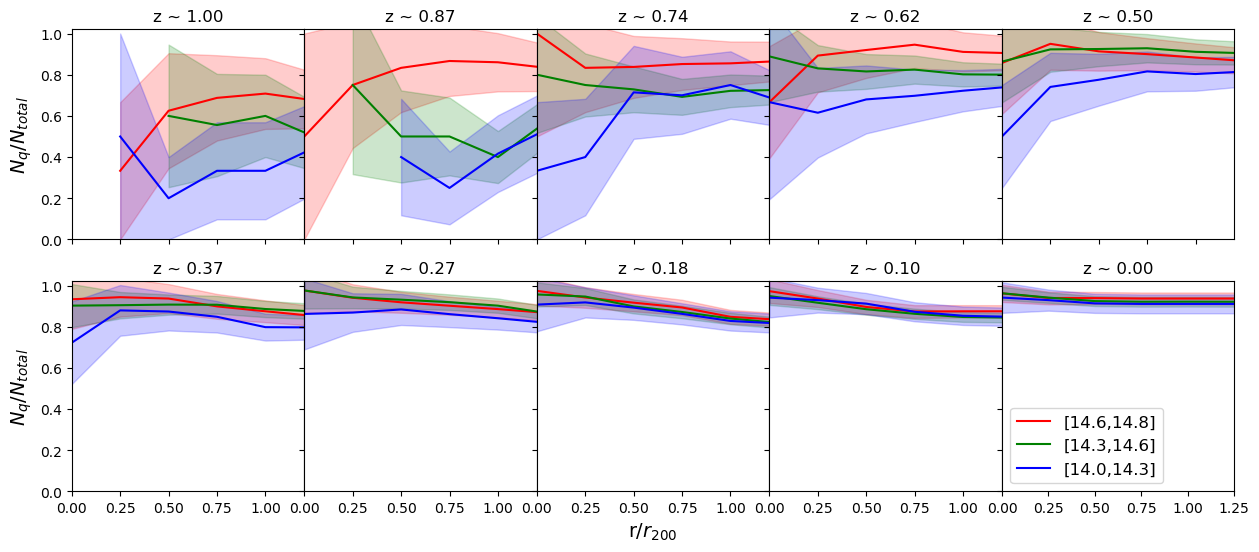}
\caption{Time evolution of the cumulative fraction of quenched galaxies within the clusters' R$_{200}$. The red, green and blue lines show the results for the high, intermediate and low mass clusters, respectively. The shaded regions show Poissonian errors.}
\label{fig:quench_r200}
\end{figure*}

It is clear from Figure~\ref{fig:fraction_strongestdrop} that centrals represent an important fraction of the pre-processed population, as they constitute $\gtrsim 50\%$ of this population in any mass bin. In isolated and low-mass galaxies several mechanisms can significantly affect the star formation history and current star formation activity. Examples are photo-reionization, which limits their gas reservoir to form stars \citep[]{Hopkins14, Chan18}, or supernova feedback which, thanks to the injection of large amounts of kinetic energy into the intergalactic medium, can eject significant fractions of the available gas \citep[]{Dekel86,Dave11,Biernacki18}. In addition, as shown by \citet{Benitez13}, ram-pressure stripping from the gas distribution within the cosmic web can efficiently remove the gas content of isolated low-mass galaxies. Pre-processing in galaxies that were not centrals at the time of $\Gamma_{\rm SFR}^{\rm SD}$ is generally associated with ram-pressure stripping within the corresponding host. However, Figure~\ref{fig:fraction_strongestdrop}
shows that the environment associated with a massive host galaxy plays a minor role 
in the pre-processing of low-mass galaxies.

As discussed before, the fraction of galaxies processed in-situ is rather low ($\lesssim 30\%$), and these  galaxies tend to be more massive than the pre-processed population at the time of their corresponding $\Gamma_{\rm SFR}^{\rm SD}$. Their most significant drop in star formation activity took place within the R$_{200}$ of the main cluster. Thus, the main mechanisms acting are tidal and ram-pressure stripping within the cluster itself. This highlights the role played by the denser environment associated with galaxy clusters.

There is a small fraction (< 30\%) of pre-processed galaxies for which $\Gamma_{\rm SFR}^{\rm SD}$ takes place in high-mass halos different from the main cluster. These halos correspond to objects that belong to massive galaxy-groups, in the mass range  $10^{13.0} \lesssim$  M$_{\rm host}$[M$_{\odot}$] $\lesssim 10^{14.0}$, that are later accreted into the main cluster.

\subsection{Critical sSFR Selection Criterion}

In Section~\ref{sec:sd} we focused on the the properties of galaxies when they suffer
their strongest drop in their star formation, $\Gamma_{\rm SFR}^{\rm SD}$. These drops do not necessarily result in the 
cessation of the star formation activity. Rather, as shown in Section~\ref{sec:sd}, on average pre-processed 
galaxies arrive 
in the cluster with a significantly lower stellar mass than those galaxies processed in-situ. Thus, 
instead of ceasing the star formation activity, an early $\Gamma_{\rm SFR}^{\rm SD}$ constrains the final galactic stellar mass.

In this Section we will focus on the moment when galaxies become effectively quenched. In  R$_{200}$ each cluster, we search for galaxies with sSFR values lower than sSFR$^{\rm Q}$ defined in 
Section~\ref{sec:def_ssfr}, and track their specific star formation history to identify the moment when this threshold is crossed. As before, we separate our galaxy sample in three bins according to cluster mass. The number of quenched galaxies in each bin is N$_{\rm gal}$ = 780, 1282 and 374 for the HMC,
IMC and LMC bins, respectively. Note that, in general, the number of quenched galaxies in each bin is $\lesssim 12\%$ smaller than the number of galaxies that have suffered some type of processing.

In the left panels of Figure~\ref{fig:fraction_wetzel} we show the host mass distribution associated with each galaxy at the time in which they became quenched. As before, for galaxies that became quenched while being centrals (green bars), M$_{\rm host} \sim $ M$_{\rm galaxy}$. Contrary to what is found with the $\Gamma_{\rm SFR}^{\rm SD}$ criterion, we find that, independently of the cluster mass bin, the vast majority of galaxies become quenched within massive hosts with 
$10^{13.5}  \lesssim $M$_{\rm host} $[M$_{\odot}]\lesssim 10^{14.5} $. This highlights the important role played by the denser 
environment of massive clusters on the overall quenching of their galaxy members. As an example we show, in Figure~\ref{fig:story_ssfr}, the time evolution of the sSFR of six galaxies in our sample as they approach 
the central galaxy of one of our clusters. The dashed lines show the time evolution of the 
clusters R$_{200}$ and the color bar the sSFR of each galaxy. The star denotes the moment when the $\Gamma_{\rm SFR}^{\rm SD}$ takes place. 
Note that galaxies in panel a) reach their quenching state as centrals. Also, it is interesting to note that the quenching state is reached as a consequence of their $\Gamma_{\rm SFR}^{\rm SD}$. For galaxies in panels b) and f), they reach their quenching state as satellites before they were accreted by the cluster, and galaxies in panels c) d) and e) are quenched inside the cluster R$_{200}$.
Also, in any case, as galaxies approach the cluster center, 
their sSFR slowly decreases. However, the change in sSFR just after the first R$_{200}$ crossing is significantly more abrupt, in some cases rapidly resulting in quenching. On the other hand, galaxies that quenched in low-mass halos, i.e. $10^{10.0}  \lesssim $M$_{\rm host}$[M$_{\odot}] \lesssim $10$^{11.0} $, did it as centrals, highlighting the regime where internal quenching processes are most relevant.  

The red bars on Figure~\ref{fig:fraction_wetzel} indicate the distributions of the in-situ quenched galaxies 
population. Interestingly, we find that the fraction of galaxies that arrived in the cluster already quenched 
(i.e., pre-quenched population) increases with cluster mass. For comparison
we find $73\%$ of the galaxies were quenched in-situ in the LMC bin, but only $45\%$ in the HMC bin. 
This apparent relation between the fraction of pre-quenched galaxies with cluster mass is further explored below. As in the case of the $\Gamma_{\rm SFR}^{\rm SD}$ criterion, we find the total mass distribution
of pre- and in-situ quenched galaxies to be very similar (medium-left panels), but they show a significant 
offset on their stellar masses at the moment of quenching (medium-right panels). 
As expected, we find that most pre-quenched galaxies ($\sim95\%$) have also been pre-processed, indicating
the important role played by the pre-processing in the quenching of low-mass objects. 
 Panel d) of Figure~\ref{fig:fraction_wetzel} 
shows the distribution of $(t_{\rm q} - t_{\rm inf})$, where $t_{\rm q}$ represents the
galaxy quenching time. We find a relation between $t_{\rm q}$ and cluster mass for the
pre-quenched population, where the high mass bin presents bigger differences between both times. 
This is a result of the hierarchical scenario; i.e., bigger clusters accrete bigger structures 
and, thus, environmental effects are more significant since earlier epochs. In general for the in-situ quenched population, we find no difference in $(t_{\rm q} - t_{\rm inf})$, 
between the different mass 
bins, highlighting the role of the virial-radius crossing in the star formation quenching of 
galaxies.

\begin{figure}
\centering
\includegraphics[width=0.5\textwidth]{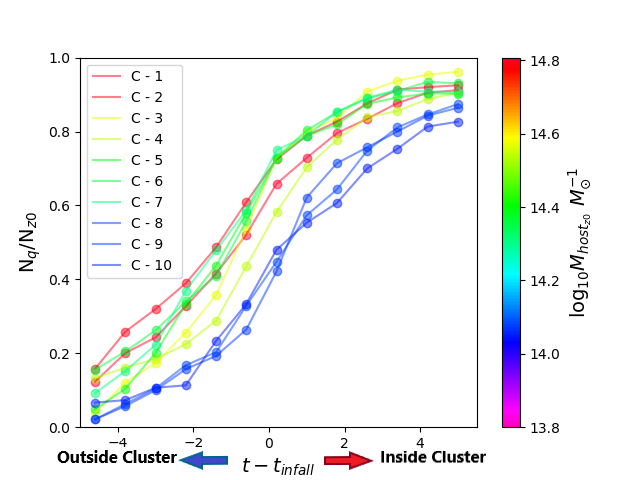}
\caption{Cumulative fraction of quenched galaxies as a function of the normalized time scale, $t-t_{\rm infall}$. The infall time $t_{\rm infall}$ is computed for each individual galaxy. The color coding indicates the total mass of each cluster at $z = 0$. Negative (positive) $t-t_{\rm infall}$ corresponds to periods of time when galaxies are located outside (inside) the cluster's $R_{200}$.}
\label{fig:t-tinf}
\end{figure}

In figure~\ref{fig:quench_r200} we show the time evolution of the cumulative fraction of quenched galaxies, 
$N_{\rm q}/N_{\rm total}$, as a function of cluster-centric distance. Here, $N_{\rm q}$ represents the number of
quenched galaxies within a given radius, $R$, and $N_{\rm total}$ the total number of galaxies within the same 
distance. The different lines correspond to the different cluster mass bins. Interestingly, we see that at early 
times, between $z \sim 1$ and $z \sim 0.5$, the fraction of quenched galaxies grows towards the cluster outskirts.
However, at later times this trend reverses, showing a decreasing fraction of quenched galaxies with distance. 
During the last decade, surveys such as WINGS \citep{Cava17} and SAMI \citep{Brough17} have shown that:
\begin{enumerate}
    \item the fraction of quenched galaxies grows towards  $z=0$. This is attributed to the environment having more time to act on cluster galaxies;
    \item the fraction of quenched galaxies decreases with cluster-centric distance. Thanks to the denser environments that can be found in the inner cluster region, galaxies, especially those with lower masses, can be more  efficiently depleted  of their gas reservoir.
\end{enumerate}
Our results are in good agreement with these observations.\\
We have previously highlighted a correlation between the fraction of pre-quenched galaxies and cluster mass. 
We further explore this correlation in Figure \ref{fig:t-tinf}. 
Here we show how the cumulative fraction of quenched galaxies,
with respect to the total number of all galaxies that can be found within R$_{200}$ at $z=0$, 
grows as a function of the normalized time, $t-t_{\rm infall}$. To obtain this plot, we first compute 
for each galaxy within  R$_{200}$ at $z=0$ the time when it crossed R$_{200}$ for the first time. 
Second, for each galaxy we define the variable
$t-t_{\rm infall}$ and identify the moment when it became quenched on this new time scale. 
Finally, we compute the cumulative quenched galaxy fraction as a function of $t-t_{\rm infall}$.  
This figure allows us to study 
  how the fraction of quenched galaxies changes as a function of the time they remain either outside 
  (negative $t-t_{\rm infall}$) or  inside (positive $t-t_{\rm infall}$) the cluster's R$_{200}$. The 
  different lines are associated with the galaxy populations of different clusters. The colors indicate the 
  mass of each cluster at $z=0$. Note that, in all clusters, the fraction of quenched galaxies slowly grows
  as galaxies approach the cluster's R$_{200}$, again highlighting the role of pre-processing. Interestingly, there is a change in the slope of this cumulative function around the time of the first R$_{200}$ crossing,
  i.e. $-1~{\rm Gyr} \lesssim t-t_{\rm infall} \lesssim 1~{\rm Gyr}$. During this
  period, the fraction of quenched galaxies raises more rapidly than during any other epoch. 
  This is in 
  agreement with the behaviour of the sSFR observed in Figure~\ref{fig:story_ssfr}, and clearly displays the role played
  by the cluster's environment. We can also observe a large dispersion in the fraction of
  galaxies that arrive quenched at the cluster's R$_{200}$, with values that go from 20 to 60$\%$. More importantly,
  this fraction shows a dependency with final cluster mass, with larger values for more massive clusters.

\begin{figure*}
\centering
\includegraphics[width=\textwidth]{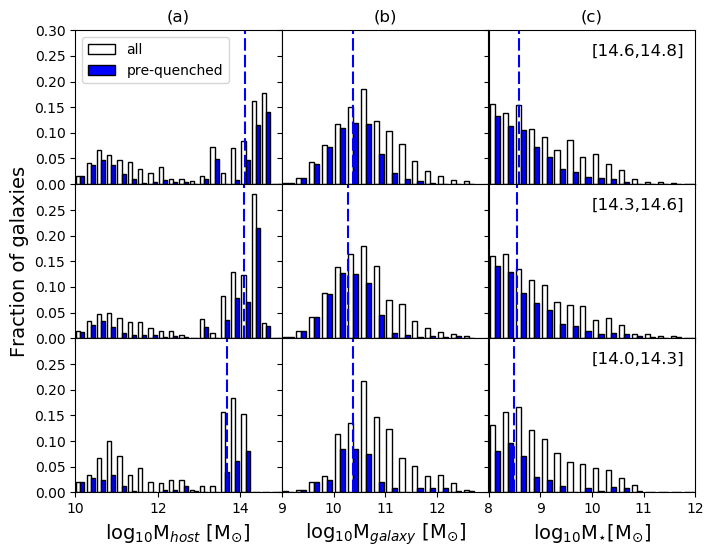}
\caption{Mass distribution of galaxies and their hosts at the moment before crossing the cluster's R$_{200}$. Each row shows the results obtained after stacking the distribution of galaxies associated with clusters within different mass ranges. Column (a) shows the mass distribution of the
host of each galaxy. Column (b) shows the distribution of galaxy total mass. Column (c) shows the distributions of stellar mass. The blue bars correspond to the galaxies quenched before the first infall and the white bars correspond to all galaxies in our sample. Dashed lines correspond to the median of the pre-quenched population.}
\label{fig:stack_halo}
\end{figure*}

To study the origin of this trend we compute the mass distribution of the structures, M$_{\rm host}$, where the quenched galaxy population at $z=0$ were located at the snapshot before their first  R$_{200}$ crossing. This is shown
in Figure~\ref{fig:stack_halo}, panels a). As before, each row corresponds to the results obtained from a 
different cluster mass bin. The blue bars indicate the fraction of pre-quenched galaxies, while the white bars show all 
the quenched galaxies found within the cluster at $z=0$. The dashed lines indicate the median for the pre-quenched population. 
Interestingly, pre-quenched galaxies on the LMC bin tend to arrive in lower mass structures than in the 
rest of the cluster mass bins. However, no significant difference is observed in both the distribution of total (M$_{\rm 
galaxy}$) and stellar masses (M$_{\star}$)
of the pre-quenched galaxy populations at infall, shown in panels b) and c), respectively. 

Our results indicate that the larger fraction of pre-quenched galaxies in larger mass clusters is the result of the hierarchical 
nature of the $\Lambda$CDM cosmological model used in this work, in which larger mass object can accrete more 
massive substructures. These more massive substructures are naturally more efficient in quenching their own galaxy
satellite population, thus resulting in a larger fraction of pre-quenched galaxies at $z=0$.

\section{Discussion and Conclusions}

In this paper we have presented a study of the different environmental-quenching and processing scenarios undergone by 
the satellite galaxies of the ten most massive clusters in the state-of-the-art EAGLE hydrodynamical simulation.
Two different criteria were defined to explore the different processes that significantly affect the SFR of 
these galaxies along their history. Our goal is to quantify and characterize the role played by the environment in these processes.

For the instantaneous strongest drop in SFR, we find that the majority of galaxies suffer their $\Gamma_{\rm SFR}^{\rm SD}$ outside the cluster's R$_{200}$  (pre-processed fraction $\gtrsim 60\%$ ). This fraction grows with cluster mass. We find that there is no correlation between the strength of the $\Gamma_{\rm SFR}^{\rm SD}$ and the time at which it occurs, nor a preferential redshift for it to happen. Nonetheless, for galaxies processed in-situ, $\Gamma_{\rm SFR}^{\rm SD}$ tends to happen at lower redshift than for the pre-processed population. 
In terms of the environment, while in-situ processing mainly occurs in massive hosts, pre-processing shows a 
strong preference to take place in galaxies that are either low mass and central ($10^{10.5}\lesssim $M$_{\rm host}$ [M$_{\odot}$]$ \lesssim 10^{11.0}$) or that belong to low-to-middle mass hosts ($10^{11}  \lesssim $ M$_{\rm host}$[M$_{\odot}]    \lesssim 10^{13.5}$ ). 
Our results are in good agreement 
with those published by \cite{Bianconi18}, who observationally studied  a sample of 23 massive clusters
(M$_{200}$ = 10$^{15.0}$[M$_{\odot}$]) with 34 infalling groups (log$_{10}$M$_{\star}$ [M$_{\odot}]$ = 10.75),
located in outer cluster regions. They found that at cluster-centric distances R$~\sim~1.3$R$_{200}$ 
the fraction of star-forming galaxies in infalling groups is half of that in the clusters. According to this,
\cite{Bianconi18} suggest that the pre-processing in groups is the responsible for these results.

Interestingly, for galaxies with similar total mass, at the time of arrival in the main cluster, the in-situ 
processed population shows in general a larger stellar mass than those pre-processed. This highlights the 
important role of pre-processing in limiting the star formation activity of low-mass galaxies. 
The origin of this pre-processing event can be explained by a variety of different internal mechanisms such as 
supernova feedback, photo-reionization, interactions and starburst phases. Unfortunately, due to the poor time and spatial 
resolution available with this simulation, it is too hard to identify what the main mechanism acting on each galaxy is. In addition, ram-pressure from the cosmic web can also cause an accelerated depletion of the gas 
reservoirs in low mass galaxies, producing abrupt changes in their star formation \citep{Benitez13}.

In the case of the Critical sSFR criterion, contrary to our results for $\Gamma_{\rm SFR}^{\rm SD}$, we find that 
quenching presents a strong preference for high-mass halos to take place. This is a strong indicator that 
dense environments promote the definitive cessation of the star formation. 

Our results are in agreement with the 
observations presented by \cite{Olave18}, who find that the fraction of high-mass (M$_{\star} \geq 10^{10.5} 
$[M$_{\odot}]$) red (i.e. passive) galaxies in clusters (i.e quenched in-situ) is higher than the fraction of high-mass red galaxies in accreted groups (i.e. pre-quenched). We find that most of the pre-quenched galaxies ($\gtrsim 
95\%$) have also been pre-processed, evidencing the importance of pre-processing in the quenching of low-mass 
galaxies. In general we find a slight preference for pre-quenching to take place at earlier times compared with quenching
in-situ. The difference in the median of the quenching time distribution is only of the order of 1 to 2 Gyr. 
As a function of cluster-centric distance, close to $z=0$ the fraction of quenched galaxies grows toward the 
cluster center. This is in good agreement with the results obtained from observational studies based on different 
surveys such as WINGS \citep{Cava17} and SAMI \citep{Brough17}. However, at earlier times, between $z \sim 1$ 
and $z \sim 0.5$, this trend reverts, showing a fraction of quenched galaxies that grows towards the cluster
outskirts. 

In general, we find that in comparison to the in-situ quenched population, on average pre-quenched galaxies have lower stellar-masses. This result appears to be in disagreement with those presented by \citet{Hou14} who found that, independent of galaxy mass, the fraction of quiescent galaxies is higher in groups than in the clusters and field. However, we can reconcile our findings with those of \citet{Hou14} by noting that those authors only studied galaxies with stellar masses in the range 9.5 $< \log_{10}$M$_{\star}[$M$_{\odot}] <$ 10.5 and with 10$^{12.0} \leq$ M$_{\rm halo}[$M$_{\odot}] \leq 10^{14.0}$. In these massive and dense substructures the environmental quenching effects are stronger.

%For $z=0$ satellite populations, we find that the fraction of quenched satellites grows as they approach the central galaxy.
We find a sharp rise in the fraction quenched satellites at the time of the first infall, highlighting the role 
played by the dense cluster environment. It is interesting to note that, although  galaxies prefer 
denser environments to reach their quenching state, the fraction of pre-quenched galaxies in our sample grows 
with the total mass of the cluster at $z= 0$. We find that $73\%$ of galaxies were quenched in-situ in the low-mass clusters,
but only  $45\%$ were quenched in-situ for the high-mass clusters. To explain why high-mass clusters show higher fractions of 
pre-quenched galaxies, we explore the mass distribution of the structures where the cluster satellite galaxies
reside at the moment of accretion. We find that high-mass clusters preferentially accrete their satellites through 
structures and groups that are significantly more massive than those accreted by low-mass clusters. This is a direct consequence 
of the hierarchical cosmological model used in these simulations. More massive clusters tend to accrete more 
massive substructures. Due to their own intracluster dense environments, these massive substructures arrive in the clusters with their satellite population already quenched.

\cite{Cora18a} explored the quenching time of galaxies, and the relevance of the environment on this process,
using the semi-analytic model SAG \citep{Cora18b}. A criterion similar to our sSFR threshold was imposed. 
According to their results, environmental effects dominate the star formation quenching of low-mass satellite 
galaxies (M$_{\star}$ < 10$^{10.1}$ [M$_{\odot}]$. These results are in good agreement with our results. Panels a) and c) of Figure~\ref{fig:fraction_wetzel} show that we also find an important fraction of low-stellar mass galaxies that are quenched within the cluster's R$_{200}$. Note that a significant fraction of the low-stellar mass galaxies that arrive in the cluster as quenched galaxies were actually quenched in the
dense environments of massive groups. This exemplifies the relevance of the environment in the quenching of the cluster satellite population.

We also find that there is a fraction of low-stellar mass galaxies that are quenched as centrals. According to \cite{Benitez13}, this can be explained through a combination of different mechanisms that are acting simultaneously on dwarf galaxies. Processes such as supernova feedback and photo-reionization can reheat the cool gas of these galaxies inducing the quenching of their star formation activity, a scenario commonly referred to as mass quenching. In addition, as previously discussed, ram-pressure stripping taking plance within the cosmic web filaments can also deplete the gas reservoir of dwarf galaxies, producing a quenching state due to the environment. 

As we mentioned before, due to the limtied number of snapshots available in the simulation, we do not have the capabilities to separate and distinguish the different overlaping processes that are influencing the star formation history of the galaxies. In a follow-up project we plan to explore these different mechanisms using more detailed hydrodynamical simulations from the C-EAGLE project. These simulations provide us with a great improvement in temporal resolution, with a temporal resolution for particles of \~125 Myr and \~ 25 Myr for three particular intervals of redshift ([0-1],[4-5],[7-8]), and 500 Myr for group catalogues  \citep[]{Barnes17,Bahe17}. Since this simulation suite also counts with a sample of 30 clusters with a M$_{200}$ in the range between of 10$^{14.0}$ < M$_{200}$[M$_{\odot}$] < 10$^{15.4}$, this study will also allow us to explore in more detail the dependency between cluster mass and fraction of pre-quenched galaxies.
%Our results also are in well agreement with previous observational works \citep[]{vandenBosch08, Peng10, Wetzel13}. 
%Diego: Si mencionamos estos articulos, y decimos que estamos en agreemeent, abria que explicar en que estamos en agreement. 

% When we compare these results with our first selection criteria, the importance of these overlaping mechanisms grows, as we can see the number of pre-processed low-stellar mass galaxies as centrals is predominant, especially for those dwarf galaxies with a stellar mass between $10^{8} <$M$_*[$M$_{\odot}]<10^{9.5}$, after this values, the central fraction of processed galaxies become negligible for both, pre-processed population and processed in-situ population. For the processed in-situ galaxies, according to the stellar-mass of these population, the environment drives the mechanism that produces the strongest drop, according to the results shows in \cite{Cora18a}.
%DIEGO:Esto no lo entendi del todo, y no me convence que haga falta

%AGREGAR DISCUSION RESPECTO AL PAPER DE WRIGHT 

\section*{Acknowledgements}
We thank Daniel Hernandez, Catalina Labayru, Ciria Lima, Antonella Monachesi and Catalina Mora for useful discussions and comments.
We also want to thank the anonymous referee for his/her many insightful comments, that greatly improved the quality of the manuscript.
We acknowledge the Virgo Consortium for making their simulation data available. The EAGLE simulations were performed using the DiRAC-2 facility at Durham, managed by the ICC, and the PRACE facility Curie based in France at TGCC, CEA, Bruyeres-le-Chatel. 
We also thank the support given by the ``Vicerrector\'ia de Investigaci\'on de la Univesidad de La Serena" for the support given by the program ``Apoyo al fortalecimiento de grupos de investigaci\'on".
DP also acknowledges financial support through the fellowship ``Becas Doctorales Institucionales ULS", granted by the ``Vicerrector\'ia de Investigaci\'on y Postgrado de la Universidad de La Serena. DP also thanks the hospitality of PUC during the stay at the university. DP and FAG acknowledge financial support from the Max Planck Society through a Partner Group grant. FAG acknowledges financial support from CONICYT through the project FONDECYT Regular Nr. 1181264. D.O-R acknowledges the financial support provided by CONICYT-PCHA through a PhD Scholarship, ``Beca Doctorado Nacional A\~{n}o 2015'', under contract 2015-21150415. P.C. acknowledges the support provided by FONDECYT postdoctoral research grant no 3160375. NP acknowlodges the support from BASAL AFB-170002 CATA, CONICYT Anillo-1477 and Fondecyt Regular 1150300. R.D. gratefully acknowledges support from the Chilean Centro de Excelencia en Astrof\'isica y Tecnolog\'ias Afines (CATA) BASAL grant
AFB-170002.

%%%%%%%%%%%%%%%%%%%%%%%%%%%%%%%%%%%%%%%%%%%%%%%%%%



\bibliographystyle{mnras}
\bibliography{quench_final} % if your bibtex file is called example.bib

%\begin{thebibliography}{99}
%\end{thebibliography}

%%%%%%%%%%%%%%%%%%%%%%%%%%%%%%%%%%%%%%%%%%%%%%%%%%

%%%%%%%%%%%%%%%%% APPENDICES %%%%%%%%%%%%%%%%%%%%%

% Don't change these lines
\bsp	% typesetting comment
\label{lastpage}
\end{document}